\documentclass{chjaa}
\usepackage{graphicx,times}
\input{epsf.sty}
\input{psfig.sty}

\begin{document}
{
   \title{Multi-parameter estimating photometric redshifts with artificial neural networks
}

   \volnopage{Vol.0 (200x) No.0, 000--000}
   \setcounter{page}{1}

   \author{Lili Li$^1$ $^2$\and Yanxia Zhang$^1$ \and Yongheng Zhao$^1$\and Dawei
   Yang$^2$}

   \institute{1. National Astronomical Observatories, Chinese Academy
of Sciences, China, 2. College of Physics Science and Information
Engineering, Hebei Normal Unversity, Shijiazhuang,\\
             \email{lily@lamost.org}        }}

   \abstract{
We calculate photometric redshifts from the Sloan Digital Sky
Survey Data Release 2 Galaxy Sample using artificial neural
networks (ANNs). Different input patterns based on various
parameters (e.g. magnitude, color index, flux information) are
explored and their performances for redshift prediction are
compared. For ANN technique, any parameter may be easily
incorporated as input, but our results indicate that using
dereddening magnitude produces photometric redshift accuracies
often better than the Petrosian magnitude or model magnitude.
Similarly, the model magnitude is also superior to Petrosian
magnitude. In addition, ANNs also show better performance when the
more effective parameters increase in the training set. Finally,
the method is tested on a sample of 79, 346 galaxies from the SDSS
DR2. When using 19 parameters based on the dereddening magnitude,
the rms error in redshift estimation is
$\sigma_\emph{z}=0.020184$. The ANN is highly competitive tool
when compared with traditional template-fitting methods where a
large and representative training set is available.
   }

   \authorrunning{ }
   \titlerunning{Multi-parameter estimating photometric redshifts }

   \maketitle

\section{Introduction}
\label{sect:intro}

Photometric redshifts refer to the redshift estimation of galaxies
using only medium- or broad-band photometry or imaging instead of
spectroscopy. There is a fact that broad band photometry is on the
order of magnitudes less time consuming than spectroscopy.
Furthermore, photometry is available for faint galaxies that are
not spectroscopically accessible, at the least, because of finite
telescope time. In addition, a greater area of the sky covered by
imaging detectors usually makes the redshifts of more objects
measured simultaneously than by spectroscopy that is only limited
to individual galaxies or those positioned on slits or fibres. The
importance of the technique is growing not only with the desire to
gain a greater understanding of galaxy evolution (for example, the
determination of luminosity function), but also in weak
gravitational lensing, where redshift estimates can reduce
contamination from intrinsic alignments (Heymans \& Heavens 2003;
King \& Schneider 2003). If the method can be found to obtain an
accurate estimate of the redshift for the larger photometric
catalog, much better constraints on the formation and evolution of
large-scale structural elements such as galaxy cluster, filaments,
and cosmological models (e.g. Blake \& Bridle 2005) in general may
be achieved. However, photometric redshifts subject to relatively
lesser precision. For many applications such as determining
properties of large numbers of galaxies and the large-scale
structure of the universe, it is quite tolerable and sometimes
even more effective.

The concept of photometric redshifts was first developed by Baum
(1962). Since then, many new methods have been applied to
calibrate photometric-redshift relations. To date, these methods
have typically been employed on multicolor deep-field and
wide-field surveys, notably the Hubble Deep Field (e.g. Gwyn \&
Hartwick 1996; Sawicki et al. 1997; Connolly et al. 1998;
Fern$\acute{a}$ndez-Soto et al. 1999; Fontana et al. 2000;
Vanzella et al. 2004; Coe et al. 2006 ) and the Sloan Digital Sky
Survey (Sowards-Emmerd et al. 2000; Casbai et al. 2003; Weinstein
et al. 2004). The most common way of estimating photometric
redshifts is the template-matching technique. It requires to
convert the photometric data of each galaxy into spectral energy
distribution (SED) and compile a library of template spectra
covering galaxy types, luminosities and redshifts in the range of
interest. For a particular target galaxy, the photometric redshift
is selected to be the redshift of the most closely matching
template spectrum; This is usually defined by minimizing the
$\chi^2$ between the template and actual magnitudes. The
template-matching photometric redshift technique makes use of the
available and reasonably detailed knowledge of galaxy SED and in
principle it may be used reliably even for populations of galaxies
with few or no spectroscopically conformed redshifts. However, its
success strongly depends on the compilation of a library of
accurate and representative template SEDs (see e.g. Hogg et al.
1998). In the situation with a large amount of prior redshift
information about the sample, the template-matching technique is
not the best approach.

An alternative approach is a polynomial or other function fitting,
mapping the photometric data to the known redshifts and using this
to estimate redshifts for the remainder of the sample with unknown
redshifts (e.g. Sowards-Emmerd et al. 1999). In essence, its aim
is to derive a parametrization for the redshift as a function of
photometric parameters. This requires a large and representative
training set of galaxies with both photometry and a precisely
known redshift. A simple example is to express the redshift as a
polynomial in the galaxy colours (Connolly et al. 1995;
Sowards-Emmerd et al. 2000). The coefficients in the polynomial
are varied to optimize the fit between the predicted and measured
redshift. The photometric redshifts for galaxies with unknown
spectroscopy can then be estimated by utilizing the optimized
function to the colours of the target galaxy.

In the recent years, a variety of techniques to estimate
photometric redshifts have emerged based on machine learning.
Artificial neural networks (ANNs) as a new possibility among the
interpolative techniques have been used in astronomy. Popular
applications include: star/galaxy separation (e.g. Odewahn \&
Nielsen 1994; Bertin \& Arnouts 1996; ), morphological
classification of galaxies (Nieversity \& Odewahn 1994; Lahav et
al. 1996; Ball et al. 2004), spectral classification (Folkes et
al. 1996; Weaver 2000). Certainly, ANNs have also been applied for
the photometric redshift prediction (Tagliaferri et al. 2002;
Firth et al. 2003; Ball et al. 2004; Vanzella et al. 2004;
Collister \& Lahav 2004).

ANNs are applicable to `mixed' data sets in which a moderately
large training set with photometry in the survey filters and
spectroscopic redshifts for the same objects are available. The
generality of using ANN method is that any parameter can be used
to train the network and make out prediction, hence photometric
redshifts can be obtained. In practice, one can measure an almost
limitless number of parameters to describe a galaxy. However, it
is desirable to have as much information as possible in the fewest
parameters, either continuous or discrete. The fewer parameters
should be physically meaningful, i.e. they should be directly
predicted by theories of galaxy and large scale structure
formation, or be related in a quantitative way. Hence, it is
necessary to find out what parameters are the most helpful and
useful for photometric redshift evaluation.

The paper explores the use of ANNs as a potential tool for
photometric redshift determination. Mainly we focus on
establishing the best set of parameters and compare the effect of
different input parameter sets on redshift estimation. The layout
of this paper is as follows. In section 2 the principle of ANNs is
introduced. Section 3 describes the data in detail and parameters
used in the experiments. The procedure to estimate redshifts with
different parameter sets is presented in Section 4. In Section 5
the performances of different parameter sets are discussed and the
conclusion is given in Section 6.

\section{Artificial Neural Networks}
\label{sect:ANNs}

ANNs, being originally conceived as models of the brain, are
collections of interconnected neurons each able to carry out
simple processing. Artificial neural networks are composed of
massively parallel distributed processors that have a natural
property for storing experiential knowledge and making it
available for use. The knowledge is acquired by the network
through a learning process and interneuron connection strengths -
known as synaptic weights - are used to store the knowledge
(Haykin 1994).

The practical applications of ANNs most often employ supervised
learning. For supervised learning, you must provide training data
that includes both the input (a set of vectors of parameters, here
each vector representing a galaxy) and the desired result or the
target value (the corresponding redshifts). After the network is
trained successfully, you can present input data alone to the ANNs
(that is, input data without the desired result), and the ANNs
will compute an output value that approximates the desired result.

This is achieved by using a training algorithm to minimize cost
function which represents the difference (error) between the
actual and desired output. The cost function $E$ is commonly of
the form

\begin{equation}
              E=\frac{1}{p}{\sum_{k=1}^p(o_k-t_k)^2}
\end{equation}
where $o_k$  and $t_k$ are the output and target respectively for
the objects. p represents the number of samples. Generally the
topology of ANNs can be schematized as a set of N layer (see
Fig.1), each layer being composed by neurons. The first layer
($i=1$ ) is usually called `input layer', the intermediate ones
`hidden layers' and the last one ($i=N$ ) `output layer'. Each
neuron $j$ in the $s$ layer derives a weighted sum of the $M$
output $z_i^{(s-1)}$ from the previous layer $(s-1)$ and, through
either a linear or a non-linear function, produces an output,

\begin{figure}
\centering
\input epsf
\epsfverbosetrue \epsfxsize 8.8cm \epsfbox{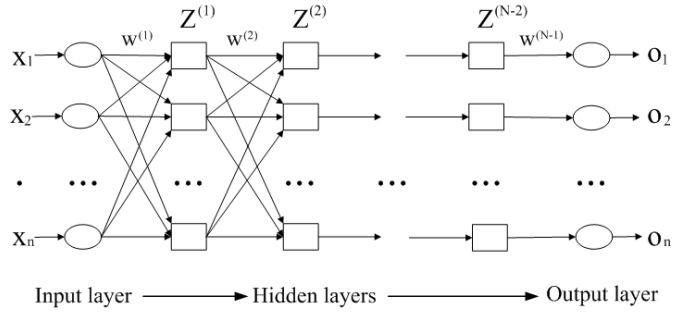} \caption{
 A schematic diagram of neural network structure. The ANNs with
 input nodes taking, for example, magnitudes in various filters,
 the middle hidden layers, and a single output node giving, for
 example, redshift $z$. Each connecting line carries a weight $w_{ij}.$
} \label{fig1}
\end{figure}

\begin{equation}
              z_j^{(s)}=f({\sum_{i=0}^M(w_{ji}^{(s)}z_i^{(s-1)}})
\end{equation}
Note that $w_{j0}$ denotes the bias for the hidden unit $j$, and
$f$ is an activation function such as the continuous sigmoid or,
as used here, the tanh function, which has an output range of -1
to 1:

\begin{equation}
              f(x)=\frac{2}{1+e^{-2x}}-1
\end{equation}

When the entire network has been executed, the outputs of the last
layer act as the output of the entire network. The free parameters
of ANNs are weighted vectors. During training neural networks, the
weights of connections are adjusted on the basis of data to
minimize the total error function. The learning produre is the so
called `back propagation'. The number of layers, the number of
neurons in each layer, and the functions are chosen from the
beginning and specify the so called `architecture' of the ANNs.

Neural networks learn by example. The neural network user gathers
representative data as a training set and initiates the weight
vector with a random seed, then invokes training algorithms to
automatically learn the structure of the data. Here we use a
popular algorithm in neural net research: the Levenberg-Marquardt
method (Levenberg 1944; Marquardt 1963, also detailed in Bishop
1995). This has the advantage that it is very quick to converge to
a minimum of the error function that may not have just a global
minimum in the multidimensional weight space but could have a
number of local minima instead. In any case, networks trained
using the exact same training set for the same number of epochs
but using different initial weights (and therefore different
starting points in this space) will converge to slightly different
final weights.

In order to avoid (possible) over-fitting during the training,
another part of the data can be reserved as a validation set
(independent both of the training and test sets, not used for
updating the weights), and used during the training to monitor the
generalization error. After a chosen number of training
iterations, training terminates and the final weights chosen for
the ANN are those from the iteration at which the cost function is
minimal on the validation set. It is called so `early stopping
method'. This is useful to avoid over-fitting to the training set
if the training set is small. But the disadvantage of this
technique is that it reduces the amount of data available for both
training and validation, which is particularly undesirable if the
data set is small.

\section{Chosen Galaxy sample and parameters}
\label{sect:sample}

The SDSS consortium has publicly released more than $10^5$
spectroscopic galaxy redshifts in the Data Release 2 (DR2). In
order to test the accuracy of the photometric redshifts derived
from SDSS DR2, we selected all objects satisfying the following
criteria (also see Vanzella et al. 2004): (1) $r$-band Petrosian
magnitude $r<17.77$; (2) the spectroscopic redshift confidence
must be greater than 0.95 and there must be no warning flags. This
gave 159 346 galaxies, which are randomly partitioned into
training, validation and test sets with respective sizes 60 000,
20 000 and 79 346. We will explore different network complexities,
the validation set is required to compare them, and the test set
is used at the end to estimate the true error of final network.

With different magnitude measurements given by SDSS, we compare
the effect of parameters for predicting redshift and give the
input patterns of many different parameter sets, which mainly
include Petrosian magnitudes, model magnitudes and dereddening
magnitudes in five different bands. The Petrosian (1976) magnitude
is based on the flux within an aperture defined by the ratio of
the local surface brightness to the mean interior surface
brightness. The model magnitude is used as a template to determine
PSF magnitude in each band. The galaxy images are fitted with the
de Vaucouleurs profile and the exponential profile of arbitrary
axis ratio and orientation. Each of these fits has a goodness and
the total magnitude associated with the better fit of the two
models is referred to as the `model' magnitude. The magnitude by
reddening correction is named dereddening magnitude. One advantage
of our ANN approach to photometric redshift estimation is that
additional parameters that can help in estimating the redshift can
be easily incorporated as extra input pattern. However, these
parameters need to be chosen carefully such that they have a
genuine dependence on the redshift. Here, we supplemented the 50\%
and 90\% Petrosian flux levels of SDSS training sample as
additional inputs to the ANNs. They are the angular radii
containing the stated fraction of the Petrosian flux. Each of
these radii is a measure of the angular size of the galaxy, which
is a redshift-dependent property.

\section{Redshift prediction with different parameter sets}
\label{sect:analysis}

The experiments were performed using ANNs in the Matlab nnet
Toolboxes. The training and test samples are independent, but in
fact it is required the training ones are representative of the
test ones. The neural network trained on the training set with
success can be applied to the test sample with the same
generalization and learning ability. During the training, we
changed a variety of net architectures and initialized the random
distributions of weights to save the `best' distribution that
corresponds to the lowest error in training sample (in almost all
cases coincident with the last epoch).

In order to evaluate the accuracy of the prediction, we define the
variance between the neural outputs (\emph{z}NN) and the targets
(spectral redshift \emph{z}spec), as below:

\begin{equation}
              \sigma_z=\sqrt{\frac{1}{N}{\sum_i(\emph{z}NN_i-\emph{z}spec_i)^2}}
\end{equation}
where N is the number of galaxies, and $i=1...N$. That is the
statistical estimate for redshift prediction of a given neural
network architecture.

\subsection{Petrosian parameters}
In this exploration, we selected the Petrosian magnitudes in five
different bands from SDSS DR2 as the root data. The Petrosian
magnitude system which measures flux in apertures is determined by
the shape of the surface brightness profile. By increasing other
parameters or changing different parameter combination as input
pattern, the neural networks were trained and the final testing
results were assessed with the Eq.4.

In this experiment, we directly used the Petrosian magnitudes in
five bands ($u, g, r, i, z$) as the first set of input parameters
for neural networks. The number of hidden units was chosen using
trial and error rather a quantitative method, such as the
 Bayesian  information criterion, because there is no one procedure
for choosing the number that applies to many datasets which is
clearly superior to trial and error. Our training therefore has
been carried out using one or two hidden layers and different
nodes. The weights corresponding to the minimum training error
have been stored. The best resulting error is
$\sigma_\emph{z}=0.027031$ with the architecture 5:10:10:1 (five
input nodes, two hidden layers with ten and ten units and one
output nodes).

To compare the effect of different parameters, we exchanged the
color index ($u-g, g-r, r-i, i-z$) and the $r$-band magnitude
($r$) as the input of neural net. Combining different
architectures with different numbers of units in the hidden layers
and selecting the best distribution of weights, the final
determined network architecture is 5:10:10:1 and the dispersion in
testing set is $\sigma_\emph{z}=0.026717$. We can see the result
is slightly better than the previous experiment. Unfortunately,
the accuracy is not adequate for our photometric redshift
estimation. So we need to consider increasing some information to
solve this problem.

Certainly, there will be more information if more parameters are
included, here the $r$-band 50 and 90 per cent Petrosian flux
radii (PetR50, PetR90) were added as two extra inputs to our ANNs.
In this case, there are 7 input parameters as input pattern. By
the experimental trial, we choose the ultimate structure is
7:16:16:1 and the rms scatter for this combination in testing set
is $\sigma_\emph{z}=0.025048$. We can see the new information
produced some improvement and 7 parameters is preferable to the
only five input patterns.

Indeed, increasing information in the training data is an obvious
method to improve the generalization. Now let us take the various
parameters into account such as the Petrosian 50 and 90 percent
flux radii in all bands and the magnitudes in the five different
bands. For this experiment, we gave 19 parameters (see Table 1.)
and the chosen networks was a single hidden layer with 20 neurons.
The experiment shows that increasing the number of nodes in the
architecture of neural network does not cause the results to
change significantly. The trial final structure is 19:20:1 and the
scatter is down to 0.021596. The result shows that correspondingly
adds some new information gives a clearly better improvement. We
compared the spectroscopic redshifts with the ANN photometric
redshifts for our experiment with 19 parameters in Fig.2. The
results of dispersion ($\sigma_\emph{z}$) for each set of
parameters to estimate photometric redshifts are summarized in
Table 1.

\begin{figure}
   \vspace{2mm}
   \begin{center}
   \hspace{3mm}\psfig{figure=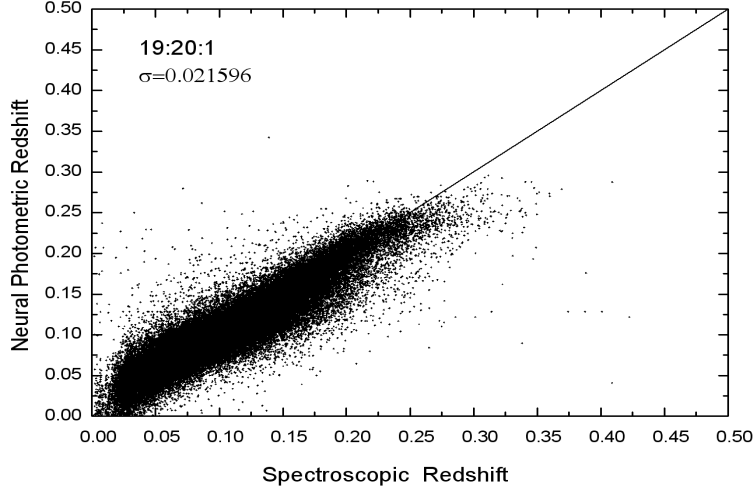,width=120mm,height=80mm,angle=0.0}
%   \parbox{180mm}{{\vspace{2mm} }}
   \caption{ the spectroscopic redshifts vs. the photometric redshifts, the redshift
prediction in the SDSS DR2 (79, 346 galaxies)
 sample using 19 Petrosian magnitude parameters.
The ANN architecture is 19:20:1 and the scatter is 0.021596   }
   \label{Fig.2}
   \end{center}
\end{figure}

\begin{table}[]
  \caption[]{ the comparison of different sets of parameters using
Petrosian magnitude  }

  \label{Tab.1}
  \begin{center}\begin{tabular}{rllllll}
  \hline\noalign{\smallskip}
\hline &Input &Parameters        &      &$\sigma_\emph{z}$     &    &$\sigma_\emph{z} $\\
       &           &            &        &(Train) & &(Test)\\
  \hline\noalign{\smallskip}
& 5  & Petro $u, g, r, i, z$             &               &0.026939   &     &0.027031\\
 & 5  & Petro $u-g, g-r, r-i, i-z, r$     &               &0.026535    &    &0.026717\\
 & 7  & Petro $u-g, g-r, r-i, i-z, r$, PetR50, PetR90  &    &0.025002   &     &0.025131\\
 & 19 & Petro $u-g, g-r, r-i, i-z, u, g, r, i, z$,\\
 &   & PetU50, PetU90, PetG50, PetG90, PetR50, &  &0.021502      &    &0.021596 \\

 &    & PetR90, PetI50, PetI90, PetZ50, PetZ90 \\
  \noalign{\smallskip}\hline
  \end{tabular}\end{center}
\end{table}

\subsection{Model parameters}
We have attempted to find the optimal set of parameters to use in
a neural network for estimating photometric redshifts, which
leaves a nonexpert asking the question `how should I decide what
parameters to use?' The comparison process can be customized by
specifying additional comparison parameter such as model
magnitude. In another experiments, we mainly focus on the model
magnitudes in five different bands or some combination with other
parameters to make a detailed comparison of different parameters
on the same sample.

Firstly, we applied the model magnitudes in five bands
($u,g,r,i,z$) as the input parameters for neural networks. A
5:12:8:1 neural network was trained for 80 epochs with validation
set leading early termination and random initialization of the
weights is adopted. The network trained by model magnitudes
produces a dispersion in testing sample $\sigma_\emph{z}=0.0233$.

Instead of using only the magnitudes, we took four model color
index ($u-g, g-r, r-i, i-z$) and the model magnitude in $r$ band
as input parameters to make comparison with the first one. In this
way, we used the same net architecture 5:12:8:1 and varied
different distribution of weights. The ultimate prediction error
at network output $\sigma_\emph{z}=0.0221$ is relatively small.

As a comparison of the parameters, we also added PetR50 and PetR90
to the above five input patterns like the Petrosian magnitude
experiment. A 7:12:8:1 network with initial 3000 epochs, has been
carried out by changing the initial random distribution of weights
and early stopping to void over-fitting during the training. Many
network runs could be used to select good and simple structure.
The final error $\sigma_\emph{z}=0.02075$ is remarkably improved.
It is comparable to the other photometric redshifts in the
literature found using neural networks, e.g. Tagliaferri et
al.(2002), Firth, Lahav \& Somerville (2003), Vanzella et
al.(2004), Collister \& Lahav (2004). So employing PetR50 and
PetR90 in this process seems to be crucial in improving the
agreement between photometric and spectroscopic redshifts.

Finally, we added some new information in the training set in
order to reduce the systematic errors. Based on the above
parameters, all the model magnitudes and the Petrosian 50 and 90
per cent flux radii in the other bands are considered and together
19 parameters (see Table 2.) are input to the network. By trial,
we selecte network architecture 19:12:8:1 and its dispersion is
$\sigma_\emph{z}=0.020465$ which is a slight improvement because
of the addition of other parameters. Fig.~3 compares the ANN
redshifts with spectroscopic redshifts for a testing set of 79,
346 galaxies with 7:12:8:1 and 19:12:8:1 networks, respectively.
In Table 2, we summarized some of the results obtained from the
above experiments.

\begin{table}[]
  \caption[]{ the comparison of different sets of parameters using
model magnitude  }

  \label{Tab.2}
  \begin{center}\begin{tabular}{rllllll}
  \hline\noalign{\smallskip}
\hline &Input &Parameters    &         &$\sigma_\emph{z} $   &       &$\sigma_\emph{z} $\\
       &           &            &        &(Train) & &(Test)\\
  \hline\noalign{\smallskip}
& 5  & model $u, g, r, i, z$             &               &0.023354   &     &0.023321\\
 & 5  & model $u-g, g-r, r-i, i-z, r$     &               &0.022006    &    &0.022097\\
 & 7  & model $u-g, g-r, r-i, i-z, r$, PetR50, PetR90  &    &0.020765   &     &0.02075\\
 & 19 & model $u-g, g-r, r-i, i-z, u, g, r, i, z$,\\
 &   & PetU50, PetU90, PetG50, PetG90, PetR50, &  &0.02034      &    &0.020465 \\

 &    & PetR90, PetI50, PetI90, PetZ50, PetZ90 \\
  \noalign{\smallskip}\hline
  \end{tabular}\end{center}
\end{table}

\begin{figure}
   \vspace{2mm}
   \begin{center}
   \hspace{3mm}\psfig{figure=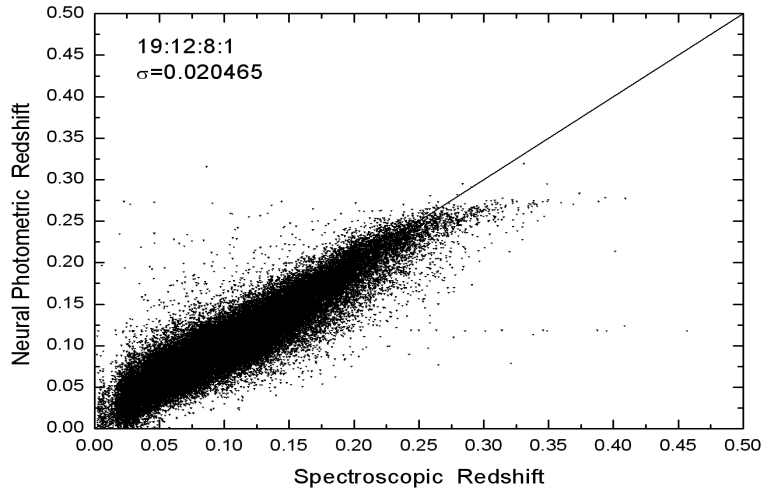,width=120mm,height=80mm,angle=0.0}
%   \parbox{180mm}{{\vspace{2mm} }}
   \caption{ A comparison of photometric and spectroscopic redshifts using
  19 model magnitude parameters. The ANN architecture
  19:12:8:1 is used. The ANNs were
 tested on a separate testing set of size 79, 346 (plotted)
 and the result is $\sigma_\emph{z}$=0.020465.  }
   \label{Fig.3}
   \end{center}
\end{figure}

\subsection{Dereddening magnitude parameters}

Similar to the procedure to predict photometric redshifts based on
petrosian and model parameters by ANNs, here we discuss the
parameter sets based on dereddening magnitudes, as well as the
 dispersion of redshift estimation. In detail, we
adopted the same sample and the training has been carried out
setting the maximum number of epochs to 3000. Different
architectures have been used with one or two hidden layers and
different numbers of nodes. For different parameter sets, the ANN
architectures are 5:5:5:1, 5:5:10:1, 7:10:1 and 19:12:1,
respectively. Correspondingly, the RMS of redshifts is listed in
Table 3. The best result of this set of parameters is
$\sigma_\emph{z}=0.020184$, where 19 parameters are considered.

Moreover, we have studied the effect of adding the error of model
magnitude (5 parameters) on redshifts estimation. Here, we give
the input patterns of network with 24 parameters, including the
above 19 dereddening magnitude and 5 model error parameters.
Finally, the network produced the scatter
$\sigma_\emph{z}=0.020053$.

\begin{table*}[ht]
\begin{center}
\caption{the comparison of different sets of parameters using
dereddening magnitude}
\bigskip
\begin{tabular}{rllllll}
\hline\hline &input &Parameters    &         &$\sigma_\emph{z} $   &       &$\sigma_\emph{z} $\\
       &           &            &        &(Train) & &(Test)\\

\hline
 & 5  & dereddening $u, g, r, i, z$             &               &0.021371   &     &0.02388\\
 & 5  & dereddening $u-g, g-r, r-i, i-z, r$     &               &0.021081    &    &0.021097\\
 & 7  & dereddening $u-g, g-r, r-i, i-z, r$, PetR50, PetR90  &    &0.020821   &     &0.020689\\
 & 19 & dereddening $u-g, g-r, r-i, i-z, u, g, r, i, z$, \\
 &    &  PetU50, PetU90, PetG50,PetG90, PetR50,  &
  &0.020174      &      &0.020184\\
 &    & PetR90, PetI50, PetI90, PetZ50, PetZ90\\

\hline

\end{tabular}
\bigskip
\end{center}
\end{table*}

\begin{figure}
   \vspace{2mm}
   \begin{center}
   \hspace{3mm}\psfig{figure=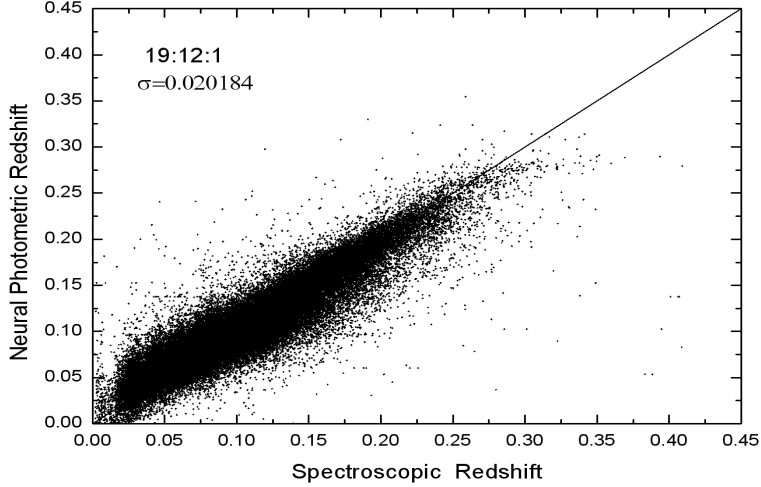,width=120mm,height=80mm,angle=0.0}
%   \parbox{180mm}{{\vspace{2mm} }}
   \caption{ Redshifts prediction using dereddening magnitude
 with 19 input parameters. The ANN architecture is
  19:12:1 and the testing sample is 79, 346 (plotted).  }
   \label{Fig.3}
   \end{center}
\end{figure}

\section{Discussion}
\label{sect:discussion}

We have presented extensive experiments with a variety of
parameters for the estimation of redshifts based on feed-forward
neural networks. There are a few things to observe about these
results. First, the experiment of Petrosian magnitudes as the root
data is listed in Table 1. The combination of four color index
with Petrosian magnitude in $r$ band ($u-g, g-r, r-i, i-z, r$) has
better performance than only five magnitudes ($u, g, r, i, z$) for
our experiment. Therefore, the second set of parameters will be
more suitable for the estimation of photometric redshifts. In
order to improve the prediction result and investigate the effect
of other parameters, we add PetR50 and PetR90 as input patterns.
The result shows that the prediction accuracy of 7 parameters
surpassed that with 5 parameters towards the same data. Finally,
as shown in Table 1 when 19 parameters are taken, the prediction
for redshifts markedly improves and the correspond system error
rate also decreases.

Secondly, we transformed the model magnitudes as basic input
pattern. It is shown in Table 2 that the second set of parameters
has yielded higher prediction accuracy than the first one, namely
the performance of the combination of model color index with the
magnitude in $r$ band is better than using only five magnitudes.
Likely, we add two other parameters (PetR50,PetR90) which will
offer some new information for training. Generally speaking, with
the increasing availability of information, the prediction should
be continually improved, because of more features considered and
more information given. Indeed, the 7 parameters concerning more
information about data have a rather good performance. We
similarly utilized 19 parameters for the estimation of photometric
redshifts and finally compared its scatter with that of the 7
parameters for the same test sample. As shown in Table 2, the
result of 19 parameters gave a slight improvement.

Thirdly, The results of using dereddening magnitudes parameters
for the sample are given in Table 3. Comparing the results,
similarly, we can see the combination of four color index with
dereddening magnitude in $r$ band ($u-g, g-r, r-i, i-z, r$) is
better than only five magnitudes ($u, g, r, i, z$). Moreover,
PetR50 and PetR90 as effective parameters also improved the
performance of neural network. When 19 parameters are considered,
more parameters giving more information, the result of prediction
for redshift is also increased.

Finally, as indicated in Table 1-3, when similarly considering the
magnitudes in five bands, the dereddening magnitudes as parameters
obtained a smallest dispersion $\sigma_\emph{z}^{test}$ among
three kinds of magnitudes and the model magnitude is better than
Petrosian magnitude. In addition, all the combinations of
parameters for dereddening magnitudes are superior to those with
Petrosian magnitudes and model magnitudes. Furthermore, there is a
slight improvement when we consider the error of model magnitude.

\section{Conclusion}
\label{sect:conclusion}

In this paper, we have described experiments comparing the
performance of a number of different parameters for estimating
photometric redshifts. From the experimental results, we can
easily see no matter using the Petrosian magnitude, the model
magnitude or the dereddening magnitude, there is a common
conclusion that the more parameters are considered, the higher the
accuracy is. As the parameters increase in the training data,
there will be more information for the network to improve its
capability of prediction and generalization, so the final accuracy
is also advanced correspondingly. Moreover, it is clear that the
performance of dereddening magnitude is superior to that using
Petrosian magnitude or model magnitude for the same parameter
structure and the same data set. Therefore, we can see the
dereddening magnitude offers some significant advantage over the
Petrosian magnitude and model magnitude, though the three sets of
parameters are available for neural networks to estimate the
photometric redshifts. Our best prediction accuracy for
photometric redshifts is $\sigma_\emph{z}=0.020184$, which is the
statistical computation of samples covered the area and which will
help large-scale structure researchers to easily study some cosmic
related issues.

With the advance in astronomical observation, there have been more
and more parameters available, it therefore becomes increasing
desirable to select the most suitable parameters among them for
astronomers to use. This is a major problem for empirical
photometric redshift estimation where inappropriate parameters
that have no obvious redshift dependence will lead to larger
scatter and error. Selecting appropriate and effective parameters
is a challenging issue in future research. In order to improve the
accuracy of estimating photometric redshifts, we will consider
more parameters from multiwavelength band, such as $J$, $H$, $K_s$
from 2MASS. Moreover we will further perform feature extraction
(e.g. principal component analysis, PCA) to reveal underlying
factors or components in a multi-dimensional parameter space.

The above neural network applications were concerned with the
photometric redshift, but neural networks have had wider
applications in astronomy. The usefulness of neural networks
derives from the fact that they are an efficient and effective
means of tackling problems which are non-linear or concerned with
multi-parameter problems. Neural network techniques for solving
problem are designed primarily to give an accurate representation
of the relationship between two sets of variables, and they are
particularly successful when the relationship is highly complex.
When implemented in estimating redshift process, it becomes
evident that neural networks are a very useful and adaptable
addition to the tools available to astronomers in tackling a wide
variety of problems (i.e. classification, regression, feature
selection).

This paper is supported by National Natural Science Foundation of
China under grant No.10473013 and No.90412016.

\label{lastpage}

\end{document}